\documentclass[prd,superscriptaddress,onecolumn,showkeys]{revtex4}
\usepackage{eurosym}
\usepackage{amsfonts}
\usepackage{array}
\usepackage{amsthm}
\usepackage{bm}
\usepackage{palatino}
\usepackage{mathpazo}
\usepackage{amssymb}
\usepackage{eurosym}
\usepackage{amsmath}
\usepackage{epsfig}
\usepackage{graphics}
\usepackage{changes}
\usepackage{color}
\usepackage{graphicx}
\usepackage[colorlinks=true,
            linkcolor=blue,
           urlcolor=black,
           citecolor=black]{hyperref}

\def\be{\begin{equation}}
\def\ee{\end{equation}}
\def\beq{\begin{eqnarray}}
\def\eeq{\end{eqnarray}}

\def\bes{\begin{eqnarray}}
\def\ees{\end{eqnarray}}

\begin{document}

\title{\textbf{Evaporation of Black Hole Under the Effect of Quantum Gravity}}

\author{Riasat Ali}
\email{riasatyasin@gmail.com}
\affiliation{Department of Mathematics, GC
University Faisalabad Layyah Campus, Layyah-31200, Pakistan}

\author{Rimsha Babar}
\email{rimsha.babar10@gmail.com}
\affiliation{Division of Science and Technology, University of Education, Vehari, Lahore-54590, Pakistan}

\author{Muhammad Asgher}
\email{m.asgher145@gmail.com}
\affiliation{Department of Mathematics, The Islamia
University of Bahawalpur, Bahawalpur-63100, Pakistan}

\author{Syed Asif Ali Shah}
\email{asifalishah695@gmail.com}
\affiliation{Department of Mathematics and Statistics, The University of Lahore 1-Km Raiwind Road,
Sultan Town Lahore 54000, Pakistan}

\begin{abstract}
This paper provides an extension for Hawking temperature of Reissner-Nordstr$\ddot{o}$m-de Sitter (RN-DS)
black hole (BH) with global monopole as well as $5$D charged black hole. We consider
the black holes metric and investigate the effects of quantum gravity ($\alpha$) on Hawking radiation. 
We investigate the charged boson particles tunneling through the
horizon of black holes by using the Hamilton-Jacobi
ansatz phenomenon. In our investigation, we study the quantum radiation to analyze the
Lagrangian wave equation with generalized uncertainty principle and calculate the
modified Hawking temperatures for black holes.
Furthermore, we analyze the charge and correction parameter
effects on the modified Hawking temperature and examine the stable and unstable condition of
RN-DS BH with global monopole as well as $5$D charged black hole.
\end{abstract}

\date{\today}
\maketitle

\section{Introduction}
Hawking examined that black hole (BH) emit thermal radiations because of the
quantum vacuum volatility effects around the event horizon \cite{1a}.
Hawking radiation can be defined by quantum tunneling method near the
horizon of BHs. The Hawking radiation was determined as the black body radiation.
This scenario indicates that the BH evaporates completely \cite{1}. The correctness
of the Hawking formula can be seen in different deductions dependent on the relativity
quantum mechanics as well as quantum theory in curved space-time \cite{2a}. An effective
model for studying the Hawking radiation is the semi-classical tunneling approach \cite{2}-\cite{15r1}.
By taking into account the different space-time in the background for this method, one
can compute the corrected Hawking temperature. The generalized BH radiate all types of
particles such as fermion, boson and scalar particles. The standard Hawking temperature
as well as corrected temperature for such types of particles in the background of various
space-time have been analyzed \cite{16a}-\cite{16c}. The corrected temperature rate is
greater than the standard temperature, which means that the different background geometries accelerates
the evaporation process.

However, the presence of a minimum observable length is predicted by different theories of quantum gravity, 
i.e., loop and string theories \cite{15a, 15b, R11!}. The generalized uncertainty Principle
(GUP) is a clear way of understanding this minimum observable length. In order to determine the GUP,
the basic commutation relations ought to be generalized. There are two different approaches to
generalize these relations, which lead to various interpretations of GUP .
Kempf et al. have put forward one way and derived the basic length \cite{15c}. Although,
Das and his collaborators re-generalized the commutation relations on the basis of the theory of doubly
special relativity as well as determined an equation of GUP \cite{17}. The maximum observable momentum
and the minimum observable length can be represented by the following expression \cite{17a}
\begin{equation}
\left[\tilde{x}_j, \tilde{p}_k\right]=\imath\hbar\left[1+\tilde{p}^2\alpha\right]\delta_{jk}.
\end{equation}
However, the GUP relation can be defined as
\begin{equation}
\Delta\tilde{x}\Delta\tilde{p}\geq\frac{\hbar}{2}\left[1+(\Delta\tilde{p})^2\alpha\right],
\end{equation}
here $\alpha=\frac{\alpha_o}{M^2_{\tilde{p}}}$. The $\alpha_o<10^{34}$ denotes the
dimensionless parameter and $M_{\tilde{p}}$ represents the Plank's mass.
Moreover, $\tilde{x}_j$, $\tilde{p}_k$ can be represented as $\tilde{x}_j
=\tilde{x}_{0j}$ and $\tilde{p}_k=\tilde{p}_{0k}(1+\tilde{p}^2\alpha)$ whereas
$\tilde{x}_{0j}$ as well as $\tilde{p}_{0k}$ agrees the relation
$\left[\tilde{x}_{0j}, \tilde{p}_{0k}\right]=\imath\hbar\delta_{jk}$.
The generalization of the basic commutation relations is not specific.
Several generalizations of commutation relations are alluded to \cite{18}-\cite{20}.
In order to obtain some knowledge about the quantum properties of gravity,
these modifications are applied extensively. Black holes are useful tools
for investigating the quantum gravity effects. Some interesting consequences
and findings were obtained by the establishment of quantum gravity effects
into BH physics by incorporating GUP \cite{21}. Furthermore, the remnant mass
and corrections to the Hawking temperature and entropy were also analyzed.
\"{O}vg\"{u}n and Jusufi \cite{22} have analyzed the GUP effects to massive
spin-$1$ and spin-$0$ particles in the background of warped DGP gravity BH.

The surface gravity and Hawking radiation by the first law of thermodynamics
through the quantum horizon has been described \cite{R4}. The results between Hawking
radiation and surface gravity of a BH have been compared. The tunneling of massive charged
particles with the emission via first law of BH thermodynamics has been analyzed \cite{R6}.
The author calculated the tunneling probability through the framework of
non-commutative quasi coordinates of Reissner-Nordstr$\ddot{o}$m (RN) BH.

Sharif and Abbas \cite{R10} analyzed the phantom energy accretion at $\mu < 0$ and formulated
equation of motion for steady state spherically symmetric spacetime. Furthermore,
they discuss about the accretion and critical accretion of BH.
They concluded that the mass of BH decreases due to phantom accretion.
Ling \cite{R11}  extended the fermion tunneling method to $5$D black lenses. As a consequence
of their analysis they argued that with the help of tunneling phenomenon one can find
correct values of Hawking temperature for static and rotating BH.
Liu et al. \cite{R17} investigated the quantum corrections, spacetime and energy conservation
of charged particles tunneling in a modified Reissner BH.
He found that the entropy does not depend on the dispersion relations of matter fields.
In \cite{R23}, the authors derive the geodesic equations of massive particles in a decent manner.
They also derive Hawking radiation via tunneling from charged black hole (CBH).
Generalized uncertainty principle effect on Hawking radiation via BH geometry
was studided by Gecim and Sucu \cite{R30}. From this analysis they
saw that the Hawking temperature increases when angular momentum increases. Shababi and Addazi \cite{z1} have analyzed the cosmological vacuum energy, potential and quantum results on new type of higher D-dimensional nonperturbative GUP models. They have \cite{z2} also discussed the  nonperturbative and quantum mechanics from GUP for D-dimension and non-commutative space. The nonlinear quantum algebra extension and functional momenta operator has been extended to the GUP and non-commutative algebra spacetime \cite{z3}. Khosropour and his colleagues \cite{z9} have also 
investigated the non-linear Schrodinger and non-linear Klein–Gordon equations in the framework of GUP. Many important
contributions have been made on extended gravity and teleparallel gravity for different BHs \cite{z4}-\cite{z7}.
The total momentum and energy of an isolated system has been  extracted in integral surface
over a closed large surface enclosing the system as well as the form of an asymptotics of parallel vector fields at distance considerable from the source and the weak gravitational field have been studied and the most general linear gravitational Lagrangian field equation has been followed in vacuum \cite{z8}.

The main objective of this paper is to provide the extension of Hawking temperature for spin-$1$
particles near the event horizon from RN-de Sitter (DS) BH with global monopole as well as $5$D CBH
by taking into account the quantum gravity effects. Firstly, we modify the Proca
equation under the influence of quantum gravity. To meet our goal, we follow the GUP framework.
In order to study the equation of motion for spin-$1$ particles, we use the Hamilton-Jacobi process.
After this, we calculate the modified tunneling probability and corrected
Hawking temperatures. Moreover, we check the stability condition of BHs via
graphical interpretation of temperature with event horizon.

We have organized our paper as follows: In Sec. \textbf{II}, we have investigated
the surface gravity and quantum gravity effect on modified tunneling probability
and corrected Hawking temperature of RN-DS BH with global monopole.
Sec. \textbf{III}, we have investigated the graphical behavior of modified temperature
w.r.t horizon to study the quantum gravity and charge effects on BH.
Sec. \textbf{IV} discuss the corrected temperature for $5$D CBH.
In Sec. \textbf{V}, we discuss the effects of charge and gravity on
modified temperature with graphs.
In the last Sec. \textbf{VI}, we have summarized our results.

\section{Ressner-Nordstr\"{o}m-De-Sitter Black Hole with Global Monopole}
The most valuable predictions of general relativity (GR) are BHs. The first nontrivial BH solution
of the Einstein field equations is called the Schwarzschild BH and its extension
with charge version is known as the RN BH. The exploration of rotating case is
familiar as Kerr and Kerr-Newman BH. After, these BHs have been described by
incorporating various sources such as acceleration, magnetic charge and
cosmological constant to the usual BH mass. The RN-DS
BH with global monopole is the most fascinating extension of the solution of Einstein field equations.

The spacetime with a global monopole parameter of the RN-DS BH can be given as \cite{25}
\begin{equation}
ds^{2}=-A(r)dt^{2}+\frac{1}{A(r)}dr^{2}+B(r)d\theta^{2}
+C(r) d\phi^{2},\label{le}
\end{equation}
where
\begin{eqnarray}
A(r)&=&1+\frac{q^2}{r^2}-\frac{2m}{r}-\left(\frac{\Lambda}{3}\right)r^2,\nonumber\\
B(r)&=&\left(1-8\pi\eta^{2}\right)r^2,\nonumber\\
C(r)&=&\left(1-8\pi\eta^{2}\right)r^2 \sin^2\theta,\nonumber
\end{eqnarray}
here $\eta$, $\Lambda$, $q$ and $m$ represents the global monopole,
cosmological constant, BH charge and mass, respectively. The BH event
horizon $r_{+}$ can be determined from setting $A(r_{+})=0$.
Furthermore, in the presence of global monopole parameter, the total mass
of Arnowitt-Deser-Misner and BH charge can be defined as \cite{16}
\begin{eqnarray}
M=m\left(1-8\pi\eta^{2}\right),~~Q=q\left(1-8\pi\eta^{2}\right).\nonumber
\end{eqnarray}
In generally, the gravity parameter is related with the property of BH stability.
The physical importance of the GUP parameter in the Lagrangian equation has been analyzed \cite{6}.
As a Lagrangian field equation, the GUP parameter is the generalization of the field equation
without singularity.
To investigate the radiation phenomenon for massive bosons, we use the
modified GUP Lagrangian equation with vector field $\chi^{\mu}$ for bosons
particles can be defined as \cite{R35, R36}
\begin{equation}
\pounds^{GUP}=\frac{1}{2}(D_{\mu}\chi_{\nu}-D_{\nu}\chi_{\mu})
(D^{\mu}\chi^{\nu}-D^{\nu}\chi^{\mu})
-\frac{1}{h}eF^{\nu\mu}\chi_{\mu}\chi_{\nu}-
\frac{m^2}{h^2}\chi_{\mu}\chi^{\nu}\nonumber\\
\end{equation}
The modified field equation can be expressed \cite{6} as follows
\begin{equation}
\partial_{\mu}(\sqrt{-g}\chi^{\nu\mu})+\sqrt{-g}\frac{m^2}{\hbar^2}\chi^{\nu}+
\sqrt{-g}\frac{i}{\hbar}eA_{\mu}\chi^{\nu\mu}
+\sqrt{-g}\frac{i}{\hbar}\chi_{\mu}eF^{\nu\mu}
+\hbar^{2}\partial_{0}\partial_{0}\partial_{0}(\sqrt{-g}g^{00}\chi^{0\nu})
-\alpha \hbar^{2}\partial_{i}\partial_{i}\partial_{i}(\sqrt{-g}g^{ii}\chi^{i\nu})=0,\label{L}
\end{equation}
here $g$ shows the determinant of coefficient matrix, $\chi^{\nu\mu}$ represents the anti-symmetric
tensor and $m$ gives the particle mass. The anti-symmetric tensor $\chi_{\nu\mu}$ can be denoted as
\begin{eqnarray}
\chi_{\nu\mu}&=&(1-\hbar^2\alpha{\partial_\nu}^2)\partial_{\nu}\chi_{\mu}-
(1-\hbar^2\alpha{\partial_\mu}^2)\partial_{\mu}\chi_{\nu}+(1-\hbar^2\alpha{\partial_\nu}^2)
\frac{i}{\hbar}eA_{\nu}\chi_{\mu}-(1-\hbar^2\alpha{\partial_\mu}^2)\frac{i}{\hbar}eA_{\mu}\chi_{\nu},\label{q}\nonumber\\
\text{and}\\
F_{\nu\mu}&=&\nabla_{\nu} A_{\mu}-\nabla_{\mu} A_{\nu},~~~\nabla_{o}=\left(1
+\hbar^2\alpha g^{00}\nabla^2_{o}\right)\nabla_{o},~~~\nabla_{i}=\left(1-\hbar^2\alpha g^{ii}\nabla^2_{i}\right)\nabla_{i},
\end{eqnarray}
here $\alpha$ is quantum gravity (GUP parameter), $A_{\mu},~e~$ and $\nabla_{\mu}$ are the BH potential,
charge of emitted particle and covariant derivative, respectively.
The elements of $\chi^{\mu}$ and $\chi^{\mu\nu}$ is defined as
\begin{eqnarray}
&&\chi^{0}=-\frac{1}{A}\chi_{0},~~~\chi^{1}=A\chi_{1},~~~\chi^{2}=\frac{1}{B}\chi_{2},~~~
\chi^{3}=\frac{1}{C}\chi_{3},~~~\chi^{01}=-\chi_{01},\nonumber\\
&&\chi^{02}=-\frac{1}{AB}\chi_{02},~~~\chi^{03}=-\frac{1}{AC}\chi_{03},~~~
\chi^{12}=\frac{A}{B}\chi_{12},~~~\chi^{13}=\frac{A}{C}\chi_{13},~~~
\chi^{23}=\frac{1}{BC}\chi_{23}.\nonumber
\end{eqnarray}
The WKB approximation can be utilized as \cite{6}
\begin{equation}
\chi_{\nu}=c_{\nu}\exp\left[\frac{i}{\hbar}S_{0}(t,r,\theta,\phi)+
\Sigma \hbar^{n}S_{n}(t,r,\theta,\phi)\right],\label{dd}
\end{equation}
here, $c_{\nu}$ and $(S_{0},~S_{n})$ are constant and arbitrary functions.

In the WKB approximation the term $\hbar$ is considered just for the $1^{st}$ order and
after neglecting the higher orders in the Lagrangian gravity Eq. (\ref{L}), we get four set of equations
\begin{eqnarray}
&&A\left[c_{1}(\partial_{0}S_{0})(\partial_{1}S_{0})+\alpha c_{1}
(\partial_{0}S_{0})^{3}(\partial_{1}S_{0})-c_{0}(\partial_{1}S_{0})^{2}
-\alpha c_{0}(\partial_{1}S_{0})^4+c_{1}eA_{0}(\partial_{1}S_{0})+c_{1}
\alpha eA_{0}(\partial_{0}S_{0})^{2}(\partial_{1}S_{0})\right]\nonumber\\
&&+\frac{1}{B}\left[c_{2}(\partial_{0}S_{0})(\partial_{2}S_{0})+\alpha c_{2}
(\partial_{0}S_{0})^3(\partial_{2}S_{0})-c_{0}(\partial_{2}S_{0})^2-\alpha c_{0}
(\partial_{2}S_{0})^4+c_{2}eA_{0}(\partial_{2}S_{0})
+c_{2}\alpha eA_{0}(\partial_{0}S_{0})^2(\partial_{2}S_{0})\right]\nonumber\\
&&+\frac{1}{C}[c_{3}(\partial_{0}S_{0})(\partial_{3}S_{0})
+\alpha c_{3}(\partial_{0}S_{0})^3(\partial_{3}S_{0})
-c_{0}(\partial_{3}S_{0})^2-\alpha c_{0}(\partial_{3}S_{0})^4
+c_{3}eA_{0}(\partial_{3}S_{0})+c_{3}eA_{0}(\partial_{0}S_{0})^2
(\partial_{3}S_{0})]\nonumber\\
&&-c_{0}m^2=0,\label{aa}
\end{eqnarray}
\begin{eqnarray}
&&-\frac{1}{A}   \left[c_{0}(\partial_{0}S_{0})(\partial_{1}S_{0})+\alpha c_{0}
(\partial_{0}S_{0})(\partial_{1}S_{0})^3-c_{1}(\partial_{0}S_{0})^{2}
-\alpha c_{1}(\partial_{0}S_{0})^{4}-c_{1}eA_{0}(\partial_{0}S_{0})-
\alpha c_{1}eA_{0}(\partial_{1}S_{0})^3\right]\nonumber\\
&&+\frac{1}{B}\left[c_{2}(\partial_{1}S_{0})(\partial_{2}S_{0})+\alpha c_{2}
(\partial_{1}S_{0})^3(\partial_{2}S_{0})-c_{1}(\partial_{2}S_{0})^{2}-\alpha
c_{1}(\partial_{2}S_{0})^{4}\right]+\frac{1}{C}\left[c_{3}(\partial_{1}S_{0})(\partial_{3}S_{0})+\alpha c_{3}
(\partial_{1}S_{0})^3(\partial_{3}S_{0})\right.\nonumber\\
&&\left.-c_{1}(\partial_{3}S_{0})^{2}-\alpha c_{1}(\partial_{3}S_{0})^{4}\right]-eA_{0}
\left[c_{0}(\partial_{1}S_{0})+\alpha c_{0}(\partial_{1}S_{0})^3
-c_{1}(\partial_{0}S_{0})-\alpha c_{1}(\partial_{0}S_{0})^3
-eA_{0}c_{1}-\alpha c_{1}eA_{0}(\partial_{0}S_{0})^{2}\right]\nonumber\\
&&-c_{1}m^2=0,\\
&&-{\frac{1}{A}}\left[c_{0}(\partial_{0}S_{0})(\partial_{2}S_{0})+\alpha c_{0}
(\partial_{0}S_{0})(\partial_{2}S_{0})^{3}-c_{2}(\partial_{0}S_{0})^{2}
-\alpha c_{2}(\partial_{0}S_{0})^4
-c_{2}eA_{0}(\partial_{0}S_{0})-\alpha c_{2}eA_{0}(\partial_{0}S_{0})^{3}\right]\nonumber\\
&&+A\left[c_{1}(\partial_{1}S_{0})(\partial_{2}S_{0})+\alpha c_{1}
(\partial_{1}S_{0})(\partial_{2}S_{0})^{3}
-c_{2}(\partial_{1}S_{0})^{2}-\alpha c_{2}(\partial_{1}S_{0})^4\right]
+\frac{1}{C}\left[c_{3}(\partial_{2}S_{0})(\partial_{3}S_{0})+\alpha c_{3}
(\partial_{2}S_{0})^{3}(\partial_{3}S_{0})\right.\nonumber\\
&&\left.-c_{2}(\partial_{3}S_{0})^{2}-\alpha c_{2}(\partial_{3}S_{0})^4\right]
-\frac{eA_{0}}{A}\left[c_{0}(\partial_{2}S_{0})+\alpha c_{0}(\partial_{2}S_{0})^3
-c_{2}(\partial_{0}S_{0})-\alpha c_{2}(\partial_{0}S_{0})^3-c_{2}eA_{0}
-\alpha eA_{0}(\partial_{0}S_{0})^2\right]\nonumber\\
&&+m^2 c_{2}=0,\\
&&-{\frac{1}{A}}\left[c_{0}(\partial_{0}S_{0})(\partial_{3}S_{0})+\alpha c_{0}
(\partial_{0}S_{0})(\partial_{3}S_{0})^{3}-c_{0}(\partial_{3}S_{0})^{2}-
c_{0}(\partial_{3}S_{0})^{4}-{eA_{0}c_3}
(\partial_{0}S_{0})-\alpha c_{3}eA_{0}(\partial_{0}S_{0})^{3}\right]\nonumber\\
&&-A\left[c_{1}(\partial_{1}S_{0})(\partial_{3}S_{0})+\alpha c_{1}
(\partial_{1}S_{0})(\partial_{3}S_{0})^{3}-c_{3}(\partial_{1}S_{0})^{2}
-\alpha c_{3}(\partial_{1}S_{0})^4\right]
+\frac{1}{B}\left[c_{2}(\partial_{2}S_{0})(\partial_{3}S_{0})+\alpha c_{2}
(\partial_{2}S_{0}){(\partial_{3}S_{0})^3}\right.\nonumber\\
&&\left.-c_{3}(\partial_{2}S_{0})^{2}-\alpha c_{3}(\partial_{2}S_{0})^4\right]-\frac{eA_{0}}{A}
\left[c_{0}(\partial_{3}S_{0})+\alpha c_{0}(\partial_{3}S_{0})^3-c_{3}
(\partial_{0}S_{0})-\alpha c_{3}(\partial_{0}S_{0})^3
-c_{3}eA_{0}-\alpha eA_{0}(\partial_{0}S_{0})^2\right]\nonumber\\
&&-m^2 c_{3}=0.\label{ab}
\end{eqnarray}
Utilizing variables separation method, we consider
\begin{equation}
S_{0}=-E_{0}t+W(r,\theta)+J\phi,\label{c1}
\end{equation}
where $E_{0}=(E-j\Omega)$, $E$ and $J$ indicate the energy and angular momentum
of particles corresponding with angle $\phi$.
After utilizing the Eq. (\ref{c1}) into Eqs. (\ref{aa})-(\ref{ab}), we get
a four by four matrix as
\begin{equation*}
F(c_{0},c_{1},c_{2},c_{3})^{T}=0.
\end{equation*}
The above matrix seems to be non-trivial. Its components
are devoted as follows
\begin{eqnarray}
F_{00}&=&-W_{r}^2-\alpha W_{r}^4-\frac{1}{AB}[W_{\theta}+\alpha W_{\theta}^4]
-\frac{1}{AC}[\dot{J}^2+\alpha\dot{J}^4]-\frac{1}{A}m^2,\nonumber\\
F_{01}&=&-[E_{0}+\alpha E_{0}^3-eA_{0}(1+\alpha E_{0}^2)]W_{r},\nonumber\\
F_{02}&=&-\frac{1}{AB}[E_{0}+\alpha E_{0}^3-eA_{0}(1+\alpha E_{0}^2)]W_{\theta},\nonumber\\
F_{03}&=&-\frac{1}{AB}[E_{0}+\alpha E_{0}^3-eA_{0}(1+\alpha E_{0}^2)]\dot{J},\nonumber\\
F_{10}&=&E_{0}W_{r}+\alpha E_{0}^3 W_{r}^4-eA_{0}(W_{r}+\alpha W_{r}^3],\nonumber\\
F_{11}&=&E_{0}^2+\alpha E_{0}^4-eA_{0}E_{0}(1+\alpha W_{r}^2)-
\frac{A}{B}(W_{\theta}^2+\alpha W_{\theta}^4)
-\frac{A}{C}[\dot{J}^2+\alpha\dot{J}^4]\nonumber\\
&-&Fm^2-eA_{0}[E_{0}+\alpha E_{0}-eA_{0}+eA_{0}\alpha W_{r}^2],\nonumber\\
F_{12}&=&\frac{A}{B}[W_{r}+\alpha W_{r}^3]W_{\theta},~~
F_{13}=\frac{A}{C}[W_{r}+\alpha W_{r}^3]W_{\theta},\nonumber\\
F_{20}&=&-\frac{1}{AB}[E_{0}W_{\theta}+\alpha E_{0}W_{\theta}^3]
-\frac{eA_{0}}{AB}[W_{\theta}+\alpha W_{\theta}],\nonumber\\
F_{21}&=&\frac{A}{B}[W_{\theta}+\alpha W_{\theta}^3]W_{r},\nonumber\\
F_{22}&=&\frac{1}{AB}[E_{0}^2+\alpha E_{0}^4-eA_{0}E_{0}-\alpha eA_{0}E_{0}^3]
-\frac{A}{B}[W_{r}^2+\alpha W_{r}^4]-\frac{1}{BC}[\dot{J}^2+\alpha\dot{J}^4]\nonumber\\
&-&\frac{1}{B}m^2-\frac{eA_{0}}{AB}[E_{0}+\alpha E_{0}^3-eA_{0}+\alpha eA_{0}E_{0}^2],\nonumber\\
F_{23}&=&\frac{1}{BC}[W_{r}+\alpha W_{r}^3]W_{\theta},\nonumber
\end{eqnarray}
\begin{eqnarray}
F_{30}&=&E_{0}[\dot{J}+\alpha \dot{J}^3]-\frac{eA_{0}}{AC}[\dot{J}+\alpha \dot{J}^3]\dot{J},\nonumber\\
F_{31}&=-&\frac{A}{C}[\dot{J}+\alpha \dot{J}^3]W_{r},~~
F_{32}=\frac{1}{BC}[\dot{J}+\alpha \dot{J}^3]W_{\theta},\nonumber\\
F_{33}&=&E_{0}^2+\alpha E_{0}^4-eA_{0}E_{0}-\alpha eA_{0}E_{0}^3+
\frac{A}{C}\left[W_{r}^2+\alpha W_{r}^4\right]-\frac{1}{BC}\left[W_{\theta}^2+\alpha W_{\theta}^4\right]\nonumber\\
&-&\frac{1}{H}m^2-\frac{eA_{0}}{AC}\left[E_{0}+\alpha E_{0}^3-eA_{0}+\alpha eA_{0}W_{\theta}^2\right],\nonumber
\end{eqnarray}
where $\dot{J}=\partial_{\phi}S_{0}$, $W_{r}=\partial_{r}{S_{0}}$ and $W_{\theta}=\partial_{\theta}{S_{0}}$.
In order to obtain the non-trivial matrix result, we put the determinant $\textbf{F}$ equal to
zero, so the action of imaginary part gets the form
\begin{equation}\label{r1}
ImW^{\pm}=\pm Im \int\sqrt{\frac{(E_{0}-eA_{0})^{2}
+X_{1}\left[1+\alpha\frac{X_{2}}{X_{1}}\right]}{F^2}}dr,
\end{equation}
where
\begin{equation}
X_{1}=-\frac{F}{H}\dot{J}^2-Fm^2,~~~
X_{2}=E_{0}^4-F^2W_{r}^4-\frac{F}{H}\dot{J}^4-(eA_{0})^2E_{0}^2-2eA_{0}E_{0}^3.\nonumber
\end{equation}
The Eq. (\ref{r1}) implies
\begin{equation}
Im W^{\pm}=\pm \pi\frac{(E_{0}-eA_{0})}{2\kappa(r_{+})}\left[1+\alpha \Xi\right],
\end{equation}
here $\Xi=6\left(m^2+\frac{W^2_{\theta}+\dot{J}^2\csc^2\theta}{r^2_+}\right)>0$
represents the kinetic energy of the emitted particles via tangent plane.

The modified tunneling rate for boson particles can be computed as
\begin{equation}
\Gamma=\frac{\Gamma_{emission}}{\Gamma_{absorption}}=\exp\left[-4Im W^+\right]=
\exp\left[{-4\pi}\frac{(E_{0}-eA_{0})}
{2\kappa(r_{+})}\right]\left[1+\alpha\Xi\right].
\end{equation}
With the help of Boltzmann factor $\Gamma_{B}=\exp\left[(E-eA_{0}/T_{H}\right]$,
the Hawking temperature under the influence of quantum gravity for RN-DS BH with
global monopole can be derived as
\begin{eqnarray}
T'_{H}&=&\left[\frac{\frac{m}{r_{+}^2}-\frac{q^2}{r_{+}^3}-\frac{\Lambda}{3}
r_{+}}{2\pi}\right]\left[1-\alpha\Xi\right].\label{F5}
\end{eqnarray}
For the massless boson particles case $(i.e., m=0)$, Eq. ({\ref{F5}}) yields
\begin{eqnarray}
T_{H}'&=&\left[\frac{-\frac{q^2}{r_{+}^3}-\frac{\Lambda}{3}r_{+}}{2\pi}\right][1-\alpha\Xi].
\end{eqnarray}
The Hawking temperature for
massive boson particle $(m\neq 0)$ is quite different from massless boson particle $(m=0)$.
For the massive boson particle, the Hawking temperature depends upon charge $q$, radius of
outer horizon $r_{+}$, cosmological constant $\Lambda$ and quantum gravity parameter $\alpha$.
We can observe that the corrected Hawking temperature does not only depend upon the BH
properties but also depends upon the mass and angular momentum
of the radiated particles and quantum corrections.
The first order correction term is same as semiclassical original Hawking term,
while the next order correction term must be smaller than the preceding term satisfying GUP.

The resulting Hawking temperature at which boson
particle tunnel out through the RN-DS BH with global monopole horizon is like to the opposite temperature
of boson particle at which they tunnel inward through the RN-DS BH with global monopole horizon ($T_{H}=-T_{H}$).
By neglecting the quantum gravity parameter $\alpha=0$, we get the temperature for
RN-DS BH with global monopole \cite{25}.
For the case $\Lambda=0$, we obtain the modified temperature for RN BH \cite{26}. Moreover,
in the absence of gravity parameter (i.e., $\alpha=0$) as well as global monopole $\eta=0$,
we recover the temperature for RN-DS BH \cite{27}. It has worth to mention here that the
global monopole does not affect the Hawking temperature.
In the absence of charge $q=0$, global monopole $\eta=0$ as well as cosmological
constant $\Lambda=0$, we get the modified temperature for Schwarzschild BH \cite{28}.
The above expression (\ref{F5}) reduces into Schwarzschild BH temperature for
$ q = 0, \alpha=0, \Lambda=0$ which reads as \cite{29}.

\section{Graphical Analysis of Temperature}
We analyze the physical importance of these graphs in the presence of
quantum gravity effects and examine the stable and unstable state of
corresponding RN-DS BH with global monopole. We investigate the graphical behavior of temperature $T_{H}'$ verses
horizon $r_{+}$ for the varying values of quantum gravity parameter $\alpha$ and charge $q$.
The Hawking temperature highly increases with the decreasing horizon $r_{+}$ and a small value of the
parameter $\Xi=1$ can make a small change in temperature, only the non-physical case describes
the instability condition of RN-DS BH with global monopole. The initial mass (larger than the remnant mass)
gives three possible condition expresses physical significance relating on different values of
horizon radius, charge and quantum gravity parameter.
\begin{itemize}
\item In FIG. \textbf{1}, the corrected temperature increases with the
increasing values of quantum gravity parameter and
Hawking temperature decreases with the increasing radius horizon
$r_{+}$ (which is physical condition) and indicates the BH stability .
\item In FIG. \textbf{2}, the corrected temperature decreases gradually when,
we increase the values of charge. The temperature exponentially decreases with
the increasing horizon such that $T_{H}'\rightarrow0$ as $r_{+}\rightarrow\infty$.
This asymptotically flat condition represents the physical and stable form of BH.
\end{itemize}
FIG. $\textbf{1}$ and $\textbf{2}$ represents the totally stable form of BH. The increase in tunneling
emission rate of RN-DS BH with global monopole with the increasing values of quantum gravity parameter is
analyzed. According to Hawking's physical phenomenon (More emission of radiations reduces
the size of BH radius). We observe this phenomenon in both plots, we observe maximum temperature
at minimum value of horizon. According GUP condition the next order corrections must be
small as compared to the standard term of Uncertainty relation. The positive temperature in
these plots also satisfies the GUP relation, when GUP conditions does not
satisfies temperature becomes negative (shows non-physical state of BH).
\begin{figure}[!tbp]
\centering
\begin{minipage}[b]{0.47\textwidth}
\includegraphics[width=\textwidth]{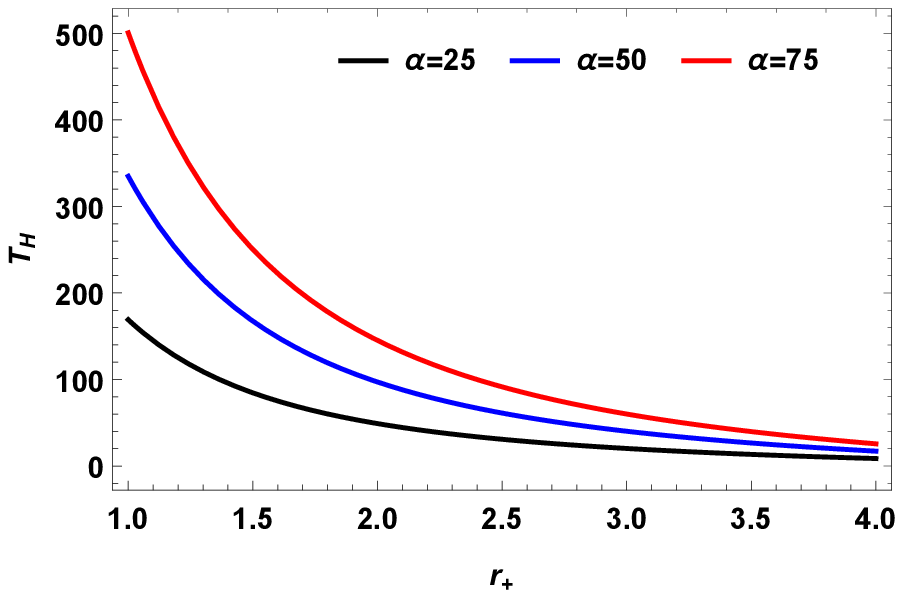}
\caption{T'$_{H}$ versus $r_{+}$ for $m =30,~q=3,~~\Lambda = 0.5$, $\Xi$=1.}
\end{minipage}
\begin{minipage}[b]{0.47\textwidth}
\includegraphics[width=\textwidth]{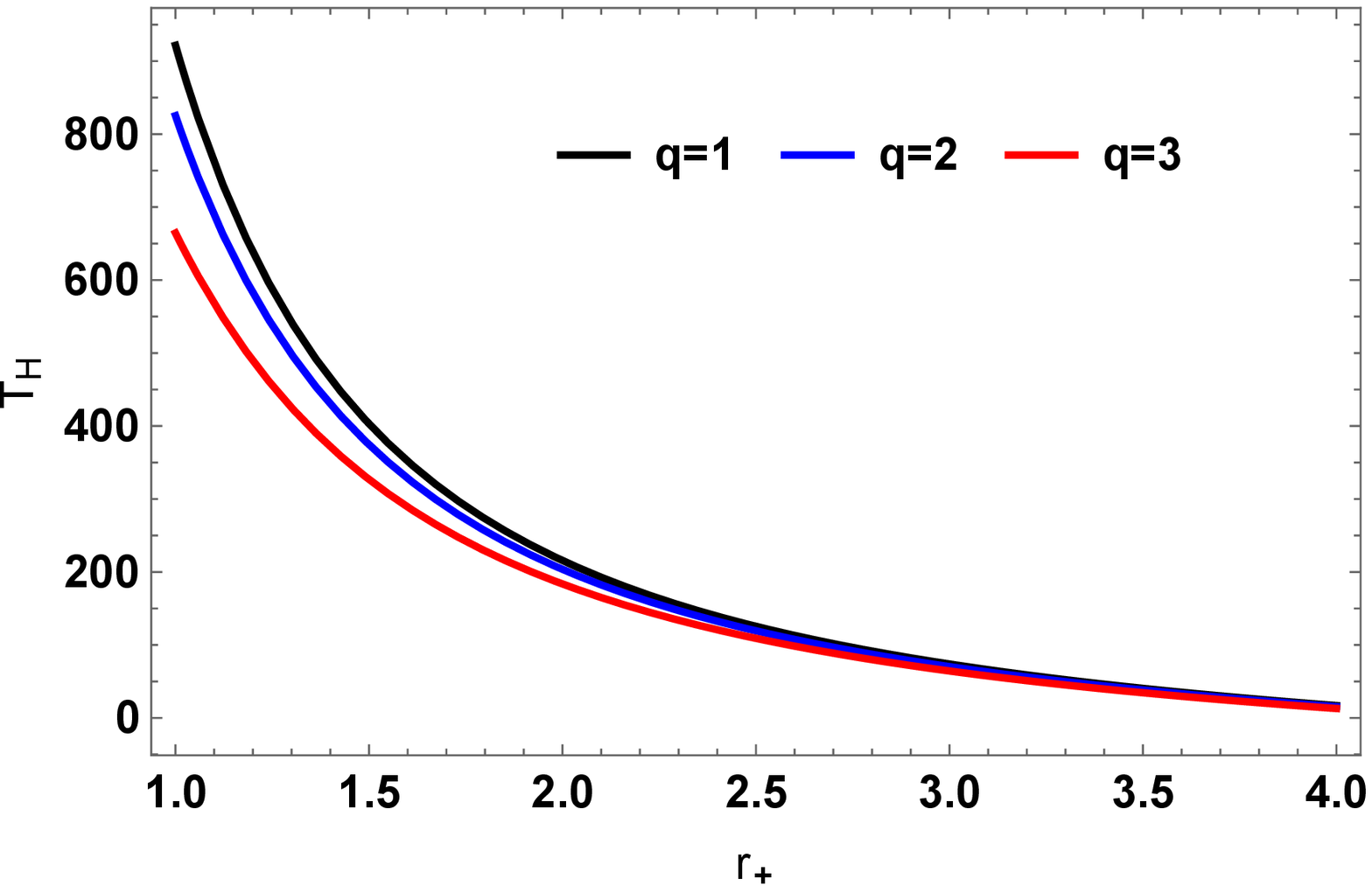}
\caption{$T'_{H}$ versus $r_{+}$ for $m =30,~~\Lambda = 0.5,
~~\alpha =100$, $\Xi=1.$}
\end{minipage}
\end{figure}

\section{$5$-Dimensional Charged Black Hole}

This section provides the extension in Hawking temperature for $5$D CBH.
We assume a charged static spherically symmetric $(n+2)$ dimensional BH solution \cite{R10},

\begin{equation}
ds^{2}=z(r)dt^2-\frac{1}{z(r)}dr^2-r^2 d\Omega_{n},
\end{equation}
where $d\Omega_{n}$ is the unit $n$ sphere and $z(r)=1+\frac{Q^2}{r^{2n-2}}-\frac{2M}{r^n-1}$.
For $n=2$ this reduce to 4D Reissner-Nordsrt\"{o}m (RN) metric while for $n=3$,
we obtain a 5D charged BH solution as
\begin{equation}
ds^{2}=\left(1-\frac{2M}{r^n-1}+\frac{Q^2}{r^{2n-2}}\right)dt^2-\left(1-\frac{2M}{r^n-1}
+\frac{Q^2}{r^{2n-2}}\right)^{-1}dr^2-r^2d\theta^2
-r^2 \sin^2\theta d\phi^2-r^2 \sin^2\theta \sin^2\phi d\psi^2,\label{A}
\end{equation}
here, $Q$ and $M$ are the charge and mass of BH respectively.

The BH horizon can be found by
solving for $r$ whose positive real roots will give horizon as follows
\begin{equation}
r_{outer}=\sqrt{M+\sqrt{M^2-Q^2}}>0,~~~~
r_{inner}=\sqrt{M-\sqrt{M^2-Q^2}}>0,\nonumber
\end{equation}
if $M^2>Q^2$, $r_{outer}>r_{inner}$ for $Q^2=M^2$, $r_{outer}=r_{inner}=\sqrt{M}$ (an extreme CBH)
and for $Q^2>M^2$, both horizons disappear and naked singularity at $r=0$ and
for $Q=0$, $r_{outer}=2\sqrt{M}$ (4D Schwarzschild) and $r_{inner}=0$. This gives that like $4$D
case, the existence of charge is necessary for the existence of Cauchy horizon (inner horizon).
The singularity of the $5$D CBH can be seen in regions,
$r_{outer}<r<\infty$, $0<r<r_{inner}$ and $r_{inner}<r<r_{outer}$.

The line element from Eq.(\ref {A}) takes the form
\begin{equation}
ds^{2}=z(r)dt^2-\frac{1}{z(r)}dr^2-r^2d^2\theta-r^2 \sin^2\theta d\phi^2
-r^2 \sin^2\theta \sin^2\phi d\psi^2.
\end{equation}

The boson particles ($W^{\pm}$ with spin equal to $1$) are significant elements for the electro-weak interaction models,
therefore the radiation of such particles should be more important in the investigation of Hawking temperature.
There are various phenomena which are analyzed for the Hawking temperature because of a spectrum of tunneling
across the horizon of $5$D CBH. The tunneling process is associated on the fundamental of physical action particles
which relates CBH radiation. The existence of positive and negative energy pairs of virtual boson particles is
like the existence of anti-particle and particle pair production. Instantaneously, the positive and negative
energy virtual boson particles are produced and annihilated in the form of particles pairs. The positive
energy virtual boson particle evaporates by the tunneling through the horizon and emit as a part of Hawking radiation.
The negative energy boson particles are immersed by the CBH or goes inside the CBH. The law of conservation of energy
and momentum are followed as in this method. The specific range of temperature at which the particles are observed stably
and unstably from the horizon of $5$D CBH. We have observed it for the $5$D CBH (which is only charged) while we have proved it
for more generalized $5$D CBH (with quantum gravity and twisting parameters etc). We know that the determination still holds
if background geometries of BH are more general.

By following the same procedure as Sec. II and utilizing Eq. (\ref{q}) for the $5$D
CBH metric, we calculate the non zero values of $\chi^{\mu}$ and $\chi^{\nu\mu}$ as follows
\begin{eqnarray}
&&\chi^{0}=\frac{\chi_{0}}{z},~~~\chi^{1}=-z\chi_{1},~~~\chi^{2}=-\frac{1}{r^2}\chi_{2},~~~
\chi^{3}=-\frac{1}{r^2 \sin^2\theta}\chi_{3},~~~~\chi^{4}=-\frac{1}{r^2 \sin^2
\theta \sin^2 \chi}\chi_{4},\chi^{01}=-\chi_{01},\nonumber\\
&&~~~\chi^{03}=-\frac{1}{zr^2 \sin^2\theta}\chi_{03},~~~\chi^{04}=-\frac{1}{zr^2 \sin^2 \theta  \sin^2\chi}\chi_{04},~~~\chi^{13}=\frac{z}{r^2 \sin^2\theta}\chi_{13},~~~
\chi^{14}=\frac{z}{r^2 \sin^2 \theta \sin^2 \chi}\chi_{14},\nonumber\\
&&
\chi^{23}=\frac{1}{r^4 \sin^2\theta}\chi_{23},\chi^{24}=\frac{1}{r^4 \sin^2 \theta \sin^2 \chi}\chi_{24},~~~
\chi^{34}=\frac{z}{r^4 \sin^4 \theta \sin^2 \chi}\chi_{34},~~~\chi^{02}=-\frac{1}{zr^2}\phi_{02},~~\chi^{12}=\frac{z}{r^2}\chi_{12}
\nonumber.
\end{eqnarray}
After putting the value of non-zero values of anti-symmetric tensor into Eq. (\ref{L})
and using WKB approximation of Eq. (\ref{dd}), we get
a set of $5$ wave equations (for simplicity,
we assume $A_{0}=A_{0}^{i}$ for all $i$)
where $c_{\nu}$ is arbitrary constant, $\Theta_{0}$ and $\Theta_{n}$ are arbitrary functions.

Using the separation of variables strategy, we have
\begin{equation}
\Theta_{0}=-(E-j_{1}\Omega_{1}-j_{2}\Omega_{2})t+W(r)+\Theta(\theta,\phi)+J \psi,
\end{equation}
where $E$ is energy of particle and $\Omega$ is angular momentum. On the other hand, $\Theta$ and $J$ denotes the
particles angular momenta associated to $\theta$, $\phi$ and $\psi$, respectively.
For the above $ \Theta_{0}$, the $4$ set of Eqs. (\ref{s1}) to (\ref{s5})
can be written in terms of $5\times5$ matrix equation.
The elements of the required matrix can be defined as
\begin{equation*}
H(c_{0},c_{1},c_{2},c_{3},c_{4})^{T}=0.
\end{equation*}
The components of above matrix are given as
are given as
\begin{eqnarray}
H_{00}&=&(\dot W^{2}+\alpha \dot W^{3})+\frac{1}{zr^2}\large[(E-j_{1}\Omega_{1}-j_{2}\Omega_{2})^{2}
+(E-j_{1}\Omega_{1}-j_{2}\Omega_{2})^{4}\large]
+\frac{1}{zr^2 \sin^2\theta}\large[\Theta^{{\prime}^{2}}+\alpha \Theta^{{\prime}^{4}}\large\large]
\nonumber\\&&+\frac{1}{zr^2 \sin^2 \theta \sin^2 \phi}\large[J^{2}+\alpha J^{4}\large]-\frac{m^2}{z},\nonumber\\
H_{01}&=&\large[(E-j_{1}\Omega_{1}-j_{2}\Omega_{2})+\alpha(E-j_{1}\Omega_{1}-j_{2}\Omega_{2})^{3}
-eA_{0}-(E-j_{1}\Omega_{1}-j_{2}\Omega_{2})^{2}\large]\dot W,\nonumber\\
H_{02}&=&\frac{1}{zr^2}[(E-j_{1}\Omega_{1}-j_{2}\Omega_{2})
+\alpha(E-j_{1}\Omega_{1}-j_{2}\Omega_{2})^{3}
-eA_{0}-\alpha eA_{0}(E-j_{1}\Omega_{1}-j_{2}\Omega_{2})]\dot\Theta,\nonumber\\
H_{03}&=&\frac{1}{zr^2 \sin^2\theta}\large[(E-j_{1}\Omega_{1}-j_{2}\Omega_{2})
+\alpha(E-j_{1}\Omega_{1}-j_{2}\Omega_{2})^3 -\alpha eA_{0}-\alpha eA_{0}(E-j_{1}
\Omega_{1}-j_{2}\Omega_{2})^{2}\large]\Theta^{\prime},\nonumber\\
H_{04}&=&\frac{1}{zr^2 \sin^2\theta \sin^2\phi}\large[(E-j_{1}\Omega_{1}-j_{2}\Omega_{2})
+\alpha(E-j_{1}\Omega_{1}-j_{2}\Omega_{2})^3
-eA_{0}-\alpha eA_{0}(E-j_{1}\Omega_{1}-j_{2}\Omega_{2})^2\large]J,\nonumber\\
H_{10}&=&(E-j_{1}\Omega_{1}-j_{2}\Omega_{2})\dot W+\alpha(E-j_{1}\Omega_{1}-j_{2}\Omega_{2})\dot W^{3}
-eA_{0}\large[\dot W+\alpha \dot W^{3}\large],\nonumber\\
H_{11}&=&(E-j_{1}\Omega_{1}-j_{2}\Omega_{2})^{2}+\alpha(E-j_{1}\Omega_{1}-j_{2}\Omega_{2})^{4}
-eA_{0}(E-j_{1}\Omega_{1}-j_{2}\Omega_{2})^{2}
-eA_{0}\alpha(E-j_{1}\Omega_{1}-j_{2}\Omega_{2})^{3}
-\frac{z}{r^2 \sin^2}\nonumber\\&&\large[\dot\Theta^{2}+\alpha \dot\Theta^{4}\large]
-\frac{1}{r^2 \sin^2 \theta}\large[\Theta^{{\prime}^{2}}+\alpha \Theta^{{\prime}^{4}}\large]
-\frac{z}{r^2 \sin^2 \theta \sin^2 \phi}\large[J^{2}+\alpha J^{4}\large]+m^2z
-eA_{0}\large[(E-j_{1}\Omega_{1}-j_{2}\Omega_{2})+\alpha\nonumber\\
&&\large[(E-j_{1}\Omega_{1}-j_{2}\Omega_{2})^3
-eA_{0}-\alpha eA_{0}(E-j_{1}\Omega_{1}-j_{2}\Omega_{2})^{2}\large],\nonumber\\
H_{12}&=&\frac{z}{r^2}\large[\dot W+\alpha \dot W^{3}\large]\dot\Theta,\nonumber\\
H_{13}&=&\frac{z}{r^2 \sin^2\theta}\large[\dot W+\alpha \dot W^{3}\large]\dot\Theta,\nonumber\\
H_{14}&=&\frac{z}{r^2 \sin^2\theta \sin^2\phi}\large[\dot W+\alpha \dot W^{3}\large]J,\nonumber\\
H_{20}&=&\frac{1}{zr^2}\large[(E-j_{1}\Omega_{1}-j_{2}\Omega_{2})\dot\Theta+\alpha(E-j_{1}\Omega_{1}
-j_{2}\Omega_{2})\dot \Theta^{3}-\frac{eA_{0}}{zr^2}\large[\dot \Theta+\alpha \dot \Theta^{3}\large],\nonumber\\
H_{21}&=&\frac{z}{r^2}\large[\dot \Theta+\alpha \dot \Theta^{3}\large]\dot W,\nonumber
\end{eqnarray}
\begin{eqnarray}
H_{22}&=&\frac{1}{zr^2}\large[(E-j_{1}\Omega_{1}-j_{2}\Omega_{2})^{2}+\alpha(E-j_{1}\Omega_{1}
-j_{2}\Omega_{2})^{4}
-eA_{0}(E-j_{1}\Omega_{1}-j_{2}\Omega_{2})-\alpha eA_{0}(E-j_{1}\Omega_{1}-j_{2}\Omega_{2})^{3}\large]\nonumber\\
&&-\frac{z}{r^2}\large[\dot W^{2}+\alpha \dot W^{4}\large]-\frac{1}{r^4 \sin^2\theta}[\Theta^{{\prime}^{2}}+
\alpha \Theta^{{\prime}^{4}}]-\frac{1}{r^4 \sin^2\theta \sin^2\phi}\large[J^2+\alpha J^4\large]+\frac{m^2}{r^2}
-\frac {eA_{0}}{zr^2}(E-j_{1}\Omega_{1}-j_{2}\Omega_{2})\nonumber\\
&&+\alpha(E-j_{1}\Omega_{1}-j_{2}\Omega_{2})^{3}-eA_{0}
-\alpha eA_{0}(E-j_{1}\Omega_{1}-j_{2}\Omega_{2})^{2}\large],\nonumber\\
H_{23}&=&\frac{1}\large[\dot \Theta+\alpha \dot \Theta^{2}\large]\Theta^{\prime},~~~
H_{24}=\frac{1}{r^2 \sin^2\theta \sin^2\phi}\large[\dot \Theta+\alpha \dot \Theta^{2}\large]J,\nonumber\\
H_{30}&=&\frac{1}{zr^2 \sin^2\theta}\large[(E-j_{1}\Omega_{1}-j_{2}\Omega_{2})\Theta^{\prime}+\alpha(E-j_{1}\Omega_{1}
-j_{2}\Omega_{2}) \Theta^{{\prime}^3}-\frac{eA_{0}}{zr^2 \sin^2\theta}\large[\Theta^\prime+\alpha\Theta^\prime\large],\nonumber\\
H_{31}&=&\frac{1}{r^2 \sin^2\theta}\large[\dot \Theta^{0}+\alpha \dot \Theta^{{\prime}^{3}}\large]\dot{W},\nonumber\\
H_{32}&=&\frac{1}{r^4 \sin^2\theta}\large[\Theta^{\prime}+\alpha \Theta^{{\prime}^{3}}\large]\dot \Theta,\nonumber\\
H_{33}&=&\frac{1}{zr^2 \sin^2\theta}\large[(E-j_{1}\Omega_{1}-j_{2}\Omega_{2})^{2}+\alpha(E-j_{1}\Omega_{1}
-j_{2}\Omega_{2})^4\large]
-\frac{1}{r^2 \sin^2\theta}\large[\dot W^{2}+\alpha \dot W^{4}\large]
-\frac{1}{r^4 \sin^2\theta}\large[\dot \Theta^{2}+\alpha \dot \Theta^{4}\large]\nonumber\\
&&
-\frac{1}{r^4 \sin^4\theta \sin^2\phi}\large[J^{2}+\alpha J^{4}\large]
+\frac{m^2}{r^2\sin^2\theta}-\frac{eA_{0}}{zr^2 \sin^2\theta}
\large[(E-j_{1}\Omega_{1}-j_{2}\Omega_{2})+\alpha(E-j_{1}\Omega_{1}
-j_{2}\Omega_{2})^3\large]-eA_{0}\nonumber\\
&-&\alpha A_{0}(E-j_{1}\Omega_{1}
-j_{2}\Omega_{2})^2\large],\nonumber\\
H_{34}&=&\frac{1}{r^4 \sin^4\theta \sin^2\theta}\large[\Theta^{\prime}+\alpha \Theta^{{\prime}^{3}}\large]J,\nonumber\\
H_{40}&=&\frac{1}{zr^2 \sin^2\theta \sin^2\phi}\large[(E-j_{1}\Omega_{1}-j_{2}\Omega_{2})J+\alpha(E-j_{1}\Omega_{1}
-j_{2}\Omega_{2})J^{3}
-\frac{eA_{0}}{zr^2 \sin^2\theta \sin^2\phi}\large[J+\alpha J^3\large],\nonumber\\
H_{41}&=&\frac{z}{r^2 \sin^2\theta \sin^2\phi}\large[(E-j_{1}\Omega_{1}-j_{2}\Omega_{2})J
+\alpha (E-j_{1}\Omega_{1}-j_{2}\Omega_{2})^2 J^{3}\large]
-\frac{eA_{0}}{r^2 \sin^2\theta \sin^2\phi}\large[J+\alpha J^{3}\large],\nonumber\\
H_{42}&=&\frac{z}{r^4 \sin^2\theta \sin^2\phi}\large[J+\alpha J^{3}\large]\dot \Theta,\nonumber\\
H_{43}&=&\frac{z}{r^4 \sin^4\theta \sin^2\phi}\large[J+\alpha  J^{3}\large]\Theta^{\prime},\nonumber
\end{eqnarray}
\begin{eqnarray}
H_{44}&=&\frac{1}{zr^2 \sin^2\theta \sin^2\phi}\large[(E-j_{1}\Omega_{1}-j_{2}\Omega_{2})^2+\alpha(E-j_{1}\Omega_{1}
-j_{2}\Omega_{2})^4-eA_{0}(E-j_{1}\Omega_{1}-j_{2}\Omega_{2})\nonumber\\&-&\alpha eA_{0}(E-j_{1}\Omega_{1}
-j_{2}\Omega_{2})^3\large]-\frac{z}{r^2 \sin^2\theta \sin^2\phi}\large[\dot W^{2}+\alpha \dot W^{4}\large]
-\frac{1}{r^4 \sin^2\theta \sin^2\phi}\large[\dot \Theta^{2}+\alpha \dot \Theta^{4}\large\large]
\nonumber\\
&-&\frac{1}{r^4 \sin^2\theta \sin^2\phi}\large[\Theta^{{\prime}^{2}}+\alpha \Theta^{{\prime}^{4}}\large]
+\frac{m^2}{r^2 \sin^2\theta \sin^2\phi}
-\frac{eA_{0}}{zr^2 \sin^2\theta \sin^2\phi}\large\nonumber\\
&&[(E-j_{1}\Omega_{1}-j_{2}\Omega_{2})+\alpha(E-j_{1}\Omega_{1}
-j_{2}\Omega_{2})^3
-eA_{0}-\alpha eA_{0}(E-j_{1}\Omega_{1}
-j_{2}\Omega_{2})^2\large],\nonumber
\end{eqnarray}
where $\dot{W}=\partial_{r}\Theta_{0}$,
$\dot{\Theta}=\partial_{\theta}\Theta_{0}$ and $\Theta^{\prime}=\partial_{\phi}\Theta_{0}$.
In order to find a non-trivial solution, we set $\big|\textbf{H}\big|=0$ and obtain
\begin{equation}\label{a1}
ImW^{\pm}=\pm \int\sqrt{\frac{(E-j_{1}\Omega_{1}-j_{2}\Omega_{2}-eA_{0})^{2}+X_{1}[1+\alpha \frac{X_{2}}{X_{1}}]}{z^2}}dr
=\pm i\pi\frac{(E-j_{1}\Omega_{1}-j_{2}\Omega_{2}-eA_{0})+[1+\Xi \alpha]}{2\kappa(r_{+})},
\end{equation}
where
\begin{eqnarray}
X_{1}&=&-\frac{\Theta^{{\prime}^{2}}}{z \sin^2\theta}-\frac{1}{z r^2 \sin^2\theta \sin^2\phi}
+\frac{m^2}{z},\nonumber\\
X_{2}&=&\frac{1}{z^2}[(E-j_{1}\Omega_{1}-j_{2}\Omega_{2})^{4}+(e{A_{0}})^2 (E-j_{1}\Omega_{1}-j_{2}\Omega_{2})^{2}
-2eA_{0}(E-j_{1}\Omega_{1}-j_{2}\Omega_{2})
-\dot W^{4}\nonumber\\&-&\frac{\Theta^{{\prime}^{4}}}{z\sin^2\theta}-\frac{J^{4}}{zr^2 \sin^2\theta \sin^2\phi}.\nonumber
\end{eqnarray}
The BH surface gravity is given by
\begin{equation}
\kappa(r_{+})=\frac{4M}{r_{+}^3}-\frac{4Q^2}{r_{+}^5}.
\end{equation}
The required boson tunneling probability can be expressed as
\begin{equation}
\Gamma=\frac{\Gamma_{emission}}{\Gamma_{absortion}}=e^{-4Im W^+}
=\exp\left[{-4\pi}\frac{(E-j_{1}\Omega_{1}-j_{2}\Omega_{2}-eA_{0})}
{\frac{4M}{{r_{+}}^3}-\frac{4Q^2}{{r_{+}}^5}}\right][1+\Xi \alpha].
\end{equation}
The corrected temperature in this case is given by
\begin{equation}
T_{H}=\frac{\frac{M}{{r_{+}}^3}-\frac{Q^2}{{r_{+}}^5}}{\pi}[1-\Xi \alpha].
\end{equation}
The Hawking temperature depends upon the mass, charge and radial coordinate $r_{+}$ of
BH. If
$\alpha$ is zero then the temperature will be independent of quantum gravity.

Stopping of the BH evaporation will be achieved in the case of quantum corrections. 
The calculations of quantum corrections slow down the increase in Hawking temperature during 
the radiation process. By continuing this procedure, there is a balance state. At this state, 
the evaporation stops and remnants are produced. It has been worth to note that if ($\alpha=0$), 
we have obtained the original Hawking temperature for $5D$ BH in the absence of quantum gravity parameter.
To estimate the residual masses of the BHs at the level of the order of the magnitude, for $Q=0$, 
the obtained temperature reduces to the case of Schwarzschild BH, i.e.,
\begin{equation}
T'_{H}=\frac{1}{8\pi M}\left[ 1-6\alpha\left(m^{2}+\frac{\left(J^{2}_{\theta}
+J^{2}_{\phi}\csc^{2}\theta\right)}{r_{+}^{2}}
\right)\right],
\end{equation}
where $\left(m^{2}+\frac{\left(J^{2}_{\theta}
+J^{2}_{\phi}\csc^{2}\theta\right)}{r_{+}^{2}}
\right)$ denotes the kinetic energy component of radiated particles
related with tangent plane at horizon. For residual mass, we approximate the kinetic energy component as $E^2$.
Quantum corrections decelerate the increase in temperature during the radiation process.
These corrections cause the radiation ceased at some specific temperature, leaving 
the remnant mass. The temperature stops increasing when this condition holds
\begin{equation}
(M-dM)(1+\alpha\Xi)\simeq M.
\end{equation}
For $dM=E$, $\alpha=\frac{\alpha_0}{M_{p}^2}$ and
$E\simeq M_{p}$, we get the following relation
\begin{equation}
M_{Res} \simeq \frac{M_{p}^2}{\alpha_0E}\gtrsim \frac{M_{p}}{\alpha_0},
~~~~~~~~~~~T_{Res}\lesssim\frac{\alpha_0}{8\pi M_p}.
\end{equation}


\section{Graphical Analysis}
We analyze the graphical expression of generalized temperature $T'_H$ versus
outer horizon $(r_{+})$ as indicated in FIG. $\textbf{3}$ and FIG. $\textbf{4}$, respectively
and for our result, we fix the arbitrary parameter $\Xi=1$ and mass $M=1000$.

\begin{itemize}
\item{In the FIG. $\textbf{3}$, the quantum gravity is fixed $\alpha=100$ and the
charge varies for different values in the range $(0\leq r_{+}\leq5)$. We can
observe that the temperature of BH decreases as we increase the values of charge. }
\item{In the FIG. $\textbf{4}$, the charge is fixed $Q=25$ and the quantum gravity varies for different
values in the range $(0\leq r_{+}\leq5)$. We can observe in plot, the temperature decreases as
we increase the values of $\alpha$.}
\end{itemize}

From both plots, we conclude that, initially, the CBH is
unstable but as the time goes on the CBH becomes stable and after a maximum height
at very high temperature, it drops down and gets an asymptotically flat condition till $r_{+}\longrightarrow \infty$.
This condition with positive temperature guarantees the BH stable condition and shows its physical form.
The negative temperature shows the unstable condition of CBH because at this point GUP condition does not satisfy.
\begin{itemize}
\item{FIG. $\textbf{5}$ indicates the behavior of temperature in the absence 
of quantum gravity parameter $\alpha=0$. We can observe that the value of temperature 
is smaller than in the presence of correction parameter as compared in the absence of 
correction parameter. One can also observe thatin the absence of gravity parameter the 
decreases with the increasing value of horizon and gets an asymptotically flat condition to ensure the BH stability.}
\end{itemize}
\begin{figure}[!tbp]
\centering
\begin{minipage}[b]{0.45\textwidth}
\includegraphics[width=\textwidth]{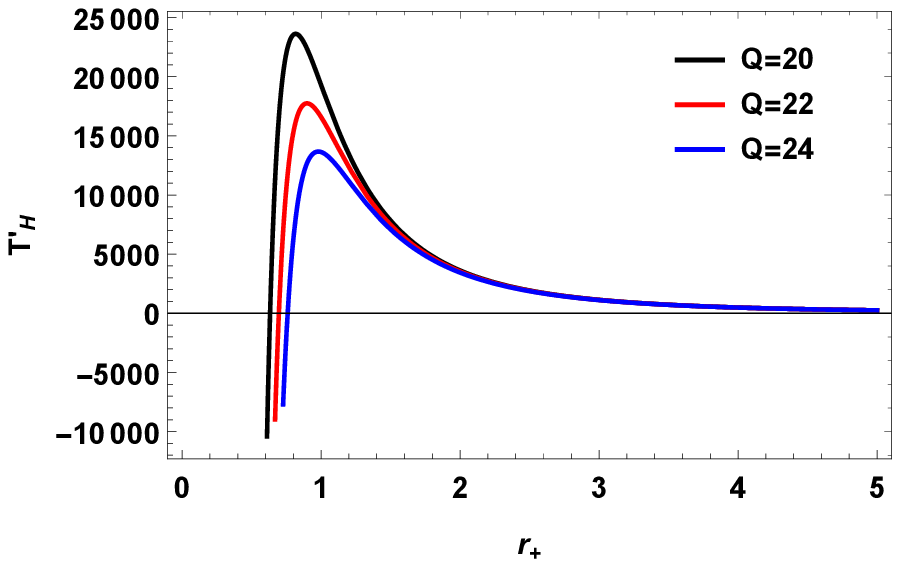}
\caption{Hawking temperature $T'_{H}$
versus outer horizon radius $r_{+}$ for $M=1000, \alpha=100$
and $\Xi=1$.}
\end{minipage}\centering
\begin{minipage}[b]{0.45\textwidth}
\includegraphics[width=\textwidth]{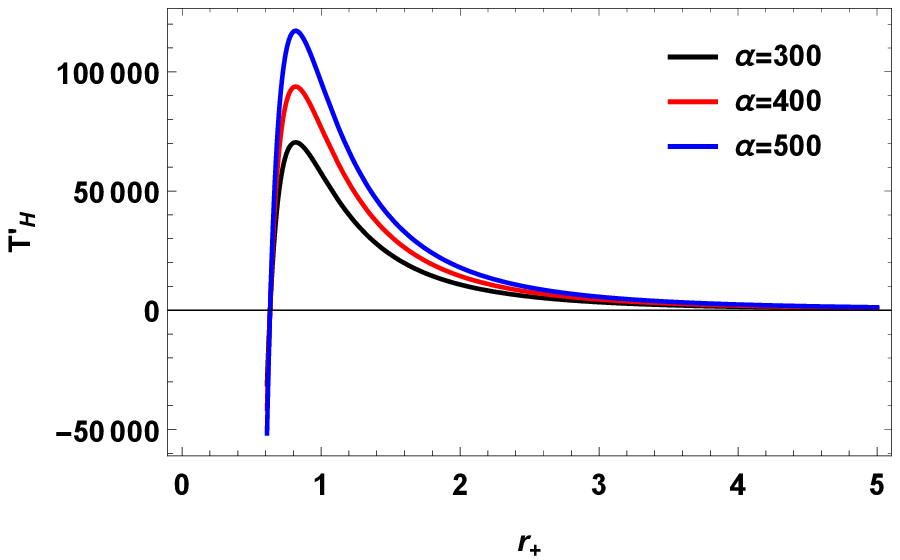}
\caption{Hawking temperature $T'_{H}$
versus outer horizon radius $r_{+}$ for $M=1000, Q=25$
and $\Xi=1$.}
\end{minipage}
\centering
\begin{minipage}[b]{0.45\textwidth}
\includegraphics[width=\textwidth]{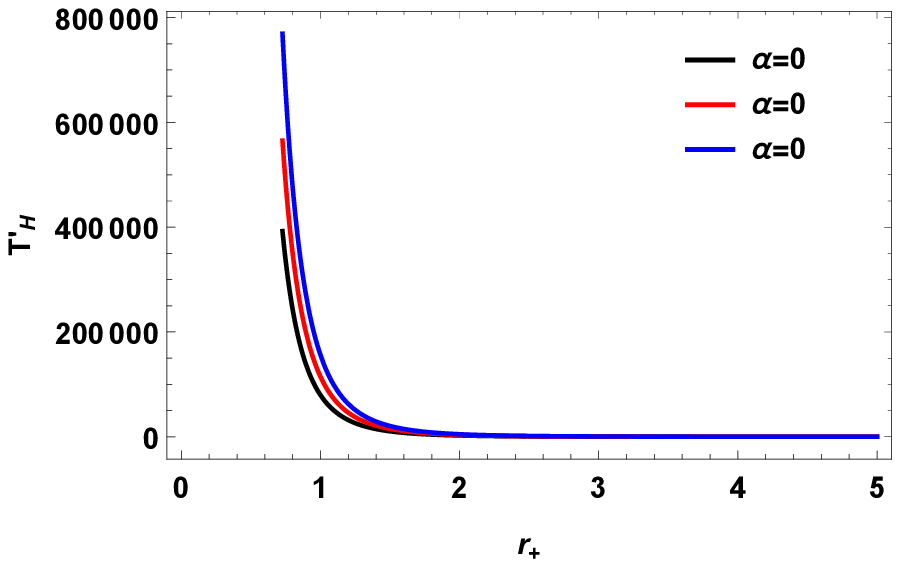}
\caption{$T'_{H}$ versus outer horizon radius $r_{+}$ for $\alpha=0$.}
\end{minipage}\centering
\end{figure}
\section{Summary and Discussion}
This paper provide an extension for Hawking temperature of RN-DS
BH with global monopole as well as $5$D CBH under the effects of quantum gravity.
In this work, we have analyzed the quantum tunneling method of massive as well as
massless boson particles from an electrically charged RN-DS
BH with global monopole. Firstly, the quantum tunneling approach has been used with $(t,~r,~\theta,~\phi)$
type of coordinate system to study the equation of motion of boson particles with GUP. For this
investigation, we have applied the Hamilton-Jacobi ansatz method and the WKB approximation
for massive charged spin-$1$ particles (bosons). The boson particle
tunneling and temperature depends upon the geometry of BH. In our phenomenon, we have modified
the Lagrangian gravity equation in curved spacetime.
We have analyzed the semiclassical action in the power series of the Planck's constant to compute
the Modified tunneling rate of the vector particles. Subsequently, the quantum
surface gravity has been computed and the back reaction effects of
the space-time has also been neglected by suitable reason of this method.
We concluded that the modified tunneling probability are not just dependent
on the BHs properties, but also on the properties of emitted vector bosons (i.e.,
particle charge, energy, surface gravity, potential and total angular momentum).
Finally, the quantum corrected Hawking temperature for corresponding BH has been derived.
Furthermore, it is important to note that the modified tunneling probability as well as the modified Hawking
temperature depend on the quantum particles, which contribute
gravitational radiation in form of massive particle
(BH's energy carrier) tunneling.
It is examined that the tunneling rate and Hawking temperature
depend on the RN-DS BH with global monopole geometry. The modified Hawking temperature depends upon
charge $q$, radius of outer horizon $r_{+}$, cosmological constant
$\Lambda$ and quantum gravity parameter $\alpha$. It is important to mention here that,
by neglecting the quantum gravity parameter $\alpha=0$, we obtain the temperature
for RN-DS BH with global monopole \cite{25}.
For the case $\Lambda=0$, we obtain the modified temperature for RN BH \cite{26}. Moreover,
in the absence of gravity parameter $\alpha=0$ as well as global monopole $\eta=0$,
we recover the temperature for RN-DS BH \cite{27}. It has worth to mention here that the
global monopole does not affect the Hawking temperature.
In the absence of charge $q=0$, global monopole $\eta=0$ as well as cosmological
constant $\Lambda=0$, we get the modified temperature for Schwarzschild BH \cite{28}.
The above expression (\ref{F5}) reduces into Schwarzschild BH temperature for
$ q = 0, \alpha=0, \Lambda=0$ which reads as \cite{29}.
Our calculations are similar to the statement that the tunneling radiation is
independent of the type of the particles and this result is also
hold for different coordinate frames by applying coordinate transformations.

Furthermore, we have investigated the physical significance of our plots. The graphical
interpretation of corrected temperature $T_{H}'$ verses horizon $r_{+}$ for different values
of gravity parameter $\alpha$ and charge $q$ depicts the
stable and unstable form of BH. From our plots, we conclude that the tunneling
emission rate increases (gives high temperature) with the increasing values
of $\alpha$. The corrected temperature
of RN-DS BH satisfies the both GUP and Hawking's conditions, that guarantee
the physical and stable states of BH.
According to Hawking's physical phenomenon (More emission of radiations reduces
the size of BH radius). We observe this phenomenon in both plots, we observe maximum temperature
at minimum value of horizon. According GUP condition the next order corrections must be
small as compared to the standard term of Uncertainty relation. The positive temperature in
these plots also satisfies the GUP relation, when GUP conditions does not
satisfies temperature becomes negative (shows non-physical state of BH).

Moreover, we have analyzed the radiation spectrum of boson particles from $5$D CBH.
By following the same process, we have calculated the corrected temperature for CBH and the corrected Hawking temperature depend
on BH mass, charge and radius of horizon, as well as on the
quantum gravity parameter of the emitted boson particles.
We have analyzed physical and non-physical behavior
of BH temperature for charge and gravity, respectively.
We have analyzed the behavior of temperature verses $r_{+}$, for
the particular ranges, the temperature decreases with the increase of gravity parameter.
When the charge was fixed and quantum gravity changes from $300$ to $500$, as well
as quantum gravity was fixed and charge change from $20$ to $24$,
we observed that the BH will be stable in the approximation range $0 \leq r+ \leq 5$
after getting a maximum height at very high temperature.
The unstable condition of BH appears at negative temperature, where GUP condition does not satisfy.
When the quantum gravity effects are ignored, we have found
the absolute Hawking temperature of $5$D CBH. Moreover, for charge free
case, the temperature and its correction correspond to the condition of Schwarzschild BH.

\section{Appendix}

The set of field equations is given as follows
\begin{eqnarray}
&&-[c_{1}(\partial_{1}\Theta_{0})(\partial_{0}\Theta_{0})+ c_{1}
(\partial_{1}\Theta_{0})(\partial_{0}\Theta_{0})^{3}\alpha-c_{0}(\partial_{1}\Theta_{0})^{2}
-(\partial_{1}\Theta_{0})^4\alpha c_{0}+(\partial_{1}\Theta_{0})c_{1}eA_{0}
+c_{1}(\partial_{1}\Theta_{0})(\partial_{0}\Theta_{0})^{2}eA_{0}]\nonumber\\
&&
-\frac{1}{zr^2}[c_{2}(\partial_{2}\Theta_{0})(\partial_{0}\Theta_{0})+\alpha
c_{2}(\partial_{2}\Theta_{0})(\partial_{0}\Theta_{0})^3
-c_{0}(\partial_{2}\Theta_{0})^2- c_{0}(\partial_{0}\Theta_{0})^4\alpha
+c_{2}(\partial_{2}\Theta_{0})eA_{0}+c_{2}(\partial_{0}
\Theta_{0})^2(\partial_{2}\Theta_{0})\alpha eA_{0}]\nonumber\\
&&
-\frac{1}{zr^2 \sin^2\theta}[c_{3}(\partial_{3}\Theta_{0})(\partial_{0}\Theta_{0})
+c_{3}(\partial_{3}\Theta_{0})(\partial_{0}\Theta_{0})^3\alpha-c_{0}
(\partial_{3}\Theta_{0})^2-c_{0}(\partial_{3}\Theta_{0})^4\alpha
+c_{3}(\partial_{3}\Theta_{0})eA_{0}+c_{3}(\partial_{3}\Theta_{0})(\partial_{0}\Theta_{0})^2eA_{0}]\nonumber\\
&&-\frac{1}{zr^2 \sin^2\theta \sin^2\phi}
[c_{4}(\partial_{4}\Theta_{0})(\partial_{0}\Theta_{0})
+c_{4}(\partial_{4}\Theta_{0})(\partial_{0}\Theta_{0})^3\alpha
-c_{0}(\partial_{4}\Theta_{0})^2
-c_{0}(\partial_{4}\Theta_{0})^4\alpha
+c_{4}(\partial_{4}\Theta_{0})eA_{0}+c_{4}eA_{0}\alpha\nonumber\\
&&(\partial_{4}\Theta_{0})(\partial_{0}\Theta_{0})^2]-\frac{m^2}{z}c_{0}=0,\label{s1}
\end{eqnarray}
\begin{eqnarray}
&&-[c_{0}(\partial_{1}\Theta_{0})(\partial_{0}\Theta_{0})+c_{0}
(\partial_{1}\Theta_{0})^3(\partial_{0}\Theta_{0})\alpha-c_{1}(\partial_{0}
\Theta_{0})^{2}-c_{1}(\partial_{0}\Theta_{0})^{4}\alpha
-c_{1}(\partial_{0}\Theta_{0})eA_{0}-c_{1}(\partial_{1}
\Theta_{0})^{3}eA_{0}\alpha]+\frac{z}{r^2}\nonumber\\
&&
[c_{2}(\partial_{2}\Theta_{0})(\partial_{1}\Theta_{0})+c_{2}
(\partial_{2}\Theta_{0})(\partial_{1}\Theta_{0})^3\alpha-c_{1}(\partial_{2}\Theta_{0})^{2}
-c_{1}(\partial_{2}\Theta_{0})^{4}\alpha]+\frac{z}{r^2 \sin^2\theta}
[c_{3}(\partial_{3}\Theta_{0})(\partial_{1}\Theta_{0})+c_{3}
(\partial_{3}\Theta_{0})(\partial_{1}\Theta_{0})^{3}\alpha\nonumber\\
&&-c_{1}(\partial_{3}\Theta_{0})^{2}-c_{1}(\partial_{3}
\Theta_{0})^{4}\alpha]+\frac{z}{r^2 \sin^2\theta \sin^2\phi}[c_{4}(\partial_{4}\Theta_{0})(\partial_{1}
\Theta_{0})+c_{4}
(\partial_{4}\Theta_{0})(\partial_{1}\Theta_{0})^{3}\alpha -c_{1}(\partial_{4}
\Theta_{0})^{2}- c_{1}(\partial_{4}\Theta_{0})^{4}\alpha]\nonumber\\
&&+m^2zc_{1}-[c_{0}(\partial_{1}\Theta_{0})+ c_{0}(\partial_{1}\Theta_{0})^{3}\alpha
-c_{1}(\partial_{0}\Theta_{0})- c_{1}(\partial_{0}\Theta_{0})^{3}\alpha
-c_{1}eA_{0}-c_{1}(\partial_{0}\Theta_{0})^{2})eA_{0}\alpha]eA_{0}=0,\label{s2}
\end{eqnarray}
\begin{eqnarray}
&&-{\frac{1}{zr^2}}[c_{0}(\partial_{2}\Theta_{0})(\partial_{0}\Theta_{0})+c_{0}
(\partial_{2}\Theta_{0})^{3}(\partial_{0}\Theta_{0})\alpha-c_{2}(\partial_{0}\Theta_{0})^{2}
-c_{2}(\partial_{0}\Theta_{0})^4\alpha -c_{2}(\partial_{0}\Theta_{0})eA_{0}
-c_{2}eA_{0}(\partial_{0}\Theta_{0})^{3}\alpha]
+\frac{z}{r^2}\nonumber\\
&&[c_{1}(\partial_{2}\Theta_{0})(\partial_{1}\Theta_{0})+c_{1}
(\partial_{2}\Theta_{0})^{3}(\partial_{1}\Theta_{0})\alpha
-c_{2}(\partial_{1}\Theta_{0})^{2}
-c_{2}(\partial_{1}\Theta_{0})^{4}\alpha]+\frac{1}{r^4 \sin^2\theta}[c_{3}(\partial_{3}\Theta_{0})(\partial_{2}\Theta_{0})+c_{3}
(\partial_{3}\Theta_{0})(\partial_{2}\Theta_{0})^{3}\alpha\nonumber\\
&&-c_{2}(\partial_{3}\Theta_{0})^{2}
-c_{2}(\partial_{3}\Theta_{0})^{4}\alpha]
+\frac{1}{r^4 \sin^2\theta \sin^2\phi}\large[c_{4}(\partial_{4}
\Theta_{0})(\partial_{2}\Theta_{0})+ c_{4}(\partial_{4}\Theta_{0})(\partial_{2}\Theta_{0})^{3}\alpha
-c_{2}(\partial_{4}\Theta_{0})^{2}
-c_{2}(\partial_{4}\Theta_{0})^{4}\alpha\large]\nonumber\\
&&+\frac{m^2 c_{2}}{r^2}-\frac{eA_{0}}{zr^2}
\large[c_{0}(\partial_{2}\Theta_{0})+ c_{0}(\partial_{2}\Theta_{0})^3\alpha-c_{2}
(\partial_{0}\Theta_{0})- c_{2}(\partial_{0}\Theta_{0})^{3}\alpha-c_{2}eA_{0}
-c_{2}(\partial_{0}\Theta_{0})^{2}\alpha eA_{0}\large]=0,\label{s3}
\end{eqnarray}
\begin{eqnarray}
&&-{\frac{1}{zr^2 sin^2\theta}}\large[c_{0}(\partial_{3}\Theta_{0})(\partial_{0}\Theta_{0})+c_{0}(\partial_{3}\Theta_{0})^{3}(\partial_{0}\Theta_{0})\alpha-c_{0}(\partial_{3}\Theta_{0})^{2}-c_{0}(\partial_{3}\Theta_{0})^{4}\alpha-c_3(\partial_{0}\Theta_{0})eA_{0}
- c_{3}(\partial_{0}\Theta_{0})^{3}\alpha eA_{0}\large]\nonumber\\
&&
+\frac{1}{r^2 \sin^2\theta}\large[c_{1}(\partial_{3}\Theta_{0})(\partial_{1}\Theta_{0})+c_{1}(\partial_{3}\Theta_{0})^{3}(\partial_{1}\Theta_{0})\alpha
-c_{3}(\partial_{1}\Theta_{0})^{2}
-c_{3}(\partial_{1}\Theta_{0})^{4}\alpha\large]+\frac{1}{r^4 \sin^2\theta}[c_{2}
(\partial_{3}\Theta_{0})(\partial_{2}\Theta_{0})+c_{2}\alpha\nonumber\\
&&
(\partial_{3}\Theta_{0})^3(\partial_{2}\Theta_{0})
-c_{3}(\partial_{2}\Theta_{0})^{2}
-c_{3}(\partial_{2}\Theta_{0})^{4}\alpha]
+\frac{1}{r^4 \sin^4\theta \sin^2\phi}\large[c_{4}
(\partial_{4}\Theta_{0})(\partial_{3}\Theta_{0})+ c_{4}(\partial_{4}\Theta_{0})(\partial_{3}\Theta_{0})^{3}\alpha
-c_{3}(\partial_{4}\Theta_{0})^{2}
\nonumber\\
&&-c_{3}(\partial_{4}\Theta_{0})^{4}\alpha\large]
+\frac{m^2 c_{3}}{r^2 \sin^2\theta}-\frac{eA_{0}}{zr^2 \sin^2\theta}
\large[c_{0}(\partial_{3}\Theta_{0})+ c_{0}(\partial_{3}\Theta_{0})^{3}\alpha
-c_{3}(\partial_{0}\Theta_{0})- c_{3}(\partial_{0}\Theta_{0})^3\alpha
-c_{3}eA_{0}-c_{3}(\partial_{0}\Theta_{0})^{2}\alpha eA_{0}\large]\nonumber\\
&&=0,\label{s4}
\end{eqnarray}
\begin{eqnarray}
&&-{\frac{1}{zr^2 \sin^2\theta \sin^2\phi}}\large[c_{0}
(\partial_{4}\Theta_{0})(\partial_{0}\Theta_{0})+c_{0}
(\partial_{4}\Theta_{0})^{3}(\partial_{0}\Theta_{0})\alpha
-c_{4}(\partial_{0}\Theta_{0})^2- c_{4}(\partial_{0}\Theta_{0})^4\alpha
-c_{4}(\partial_{0}\Theta_{0})eA_{0}-
c_{4}(\partial_{0}\Theta_{0})^{3}eA_{0}\alpha\large]\nonumber\\
&&
+\frac{z}{r^4 \sin^2\theta \sin^2\phi}\large[c_{1}(\partial_{4}\Theta_{0})(\partial_{1}\Theta_{0})+c_{1}(\partial_{4}\Theta_{0})^{3}(\partial_{1}\Theta_{0})\alpha-c_{4}(\partial_{1}\Theta_{0})^{2}
-c_{4}(\partial_{1}\Theta_{0})^{4}\alpha\large]+\frac{1}{r^4 \sin^2\theta \sin^2\phi}\nonumber\\
&&\large[c_{2}(\partial_{4}\Theta_{0})(\partial_{2}\Theta_{0})+ c_{4}(\partial_{4}\Theta_{0})^{3}(\partial_{2}\Theta_{0})\alpha-c_{4}(\partial_{2}\Theta_{0})^{2}
-c_{4}(\partial_{2}\Theta_{0})^{4}\alpha\large]+\frac{1}{r^4 \sin^4\theta \sin^2\phi}
\large[c_{3}(\partial_{4}\Theta_{0})(\partial_{3}\Theta_{0})\nonumber\\
&&+c_{4}(\partial_{4}\Theta_{0})^{3}(\partial_{3}\Theta_{0})\alpha-c_{4}(\partial_{3}\Theta_{0})^{2}
-c_{4}(\partial_{3}\Theta_{0})^{4}\alpha\large]
+\frac{m^2 c_{4}}{r^2 \sin^2\theta \sin^2\phi}-\frac{eA_{0}}{zr^2 \sin^2\theta \sin^2\phi}
\large[c_{0}(\partial_{4}\Theta_{0})+ c_{0}(\partial_{4}\Theta_{0})^{3}\alpha
\nonumber\\
&&-c_{4}
(\partial_{0}\Theta_{0})- c_{4}(\partial_{0}\Theta_{0})^{3}\alpha-c_{4}eA_{0}
-c_{4}(\partial_{0}\Theta_{0})^{2}eA_{0}\alpha\large]=0.\label{s5}
\end{eqnarray}

\end{document}